\documentclass[reprint, prb]{revtex4-1}
	\usepackage{graphicx}
	\usepackage{amsmath}
	\usepackage{amsfonts}
	\usepackage{amssymb}
	\usepackage{color}
	\usepackage[colorlinks=true, citecolor=blue, urlcolor=blue, linkcolor=black]{hyperref}

\begin{document}
	\title{Probing superfluid \(^4\mathbf{He}\) with high-frequency nanomechanical resonators down to \(\mathbf{mK}\) temperatures}
	\date{\today}

	\author{A.\,M. ~Gu\'{e}nault}
	\author{A.~Guthrie}\email{a.guthrie1@lancaster.ac.uk}
	\author{R.\,P.~Haley}
	\author{S.~Kafanov}\email{sergey.kafanov@lancaster.ac.uk}
	\author{Yu.\,A.~Pashkin}
	\author{G.\,R.~Pickett}
	\author{M.~Poole}
	\author{R.~Schanen}
	\author{V.~Tsepelin}
	\author{D.\,E. ~Zmeev}

	\affiliation{Department of Physics, Lancaster University, Lancaster, LA1 4YB, United Kingdom}

	\author{E.~Collin}
	\author{O.~Maillet}
	\author{R.~Gazizulin}
	\affiliation{Universit\'{e} Grenoble Alpes, CNRS Institut N\'{E}EL, BP 166, 38042, Grenoble Cedex 9, France}

	\begin{abstract}
		Superfluids, such as superfluid \(^3\mathrm{He}\) and \(^4\mathrm{He}\), exhibit a broad range of quantum phenomena and excitations which are unique to these systems. Nanoscale mechanical resonators are sensitive and versatile force detectors with the ability to operate over many orders of magnitude in damping. Using nanomechanical-doubly clamped beams of extremely high quality factors ($Q>10^6$), we probe superfluid \(^4\mathrm{He}\) from the superfluid transition temperature down to \(\mathrm{mK}\) temperatures at frequencies up to $11.6 \, \mathrm{MHz}$. Our studies show that nanobeam damping is dominated by hydrodynamic viscosity of the normal component of \(^4\mathrm{He}\) above \(1\,\mathrm{K}\). In the temperature range  \(0.3-0.8\,\mathrm{K}\), the ballistic quasiparticles (phonons and rotons) determine the beams' behavior. At lower temperatures, damping saturates and is determined either by magnetomotive losses or acoustic emission into helium. It is remarkable that all these distinct regimes can be extracted with just a single device, despite damping changing over six orders of magnitude.
	\end{abstract}

	\maketitle

	Mechanical resonators are frequently used as tools in experiments involving quantum fluids such as: investigation of quantum turbulence \cite{blavzkova2007transition,sheshin2008characteristics, bradley2009transition, blavzkova2009generation}, cavitation \cite{blavzkova2008cavitation, blavzkova2008cavitation1}, acoustics in superfluids \cite{bradley2012crossover, salmela2011acoustic, schmoranzer2011acoustic}, for thermometry at low-temperatures \cite{Blaauwgeers2007quartz}, and quasiparticle detection in superfluid \(^3\mathrm{He}\) \cite{bradley2017andreev}. Nanoelectromechanical systems (NEMS) are increasingly attracting interest from the scientific community due to their small size, high intrinsic quality factors and exceptional force sensitivity \cite{ekinci2005nanoelectromechanical,defoort2014slippage}. Such unique characteristics make NEMS the perfect candidates for sensing applications in modern low temperatures experiments, such as probing the complex properties of the quantum fluids \(^3\mathrm{He}\) and \(^4\mathrm{He}\). NEMS provide a unique opportunity to probe superfluids at lengths comparable to the de-Broglie wavelength of thermal excitations in superfluid \(^4\mathrm{He}\), and the quantum coherence length in superfluid \(^3\mathrm{He}\). Additionally, NEMS have demonstrated unparalleled mass sensitivity on the order of several yoctograms \cite{chaste2012nano, burg2007weighing}, which could be utilized for the detection of quantum vortices in superfluids \cite{kamppinen2018nanomechanical}.  
	\begin{figure}[b]
		\centering   
		\includegraphics[width=0.98 \linewidth]{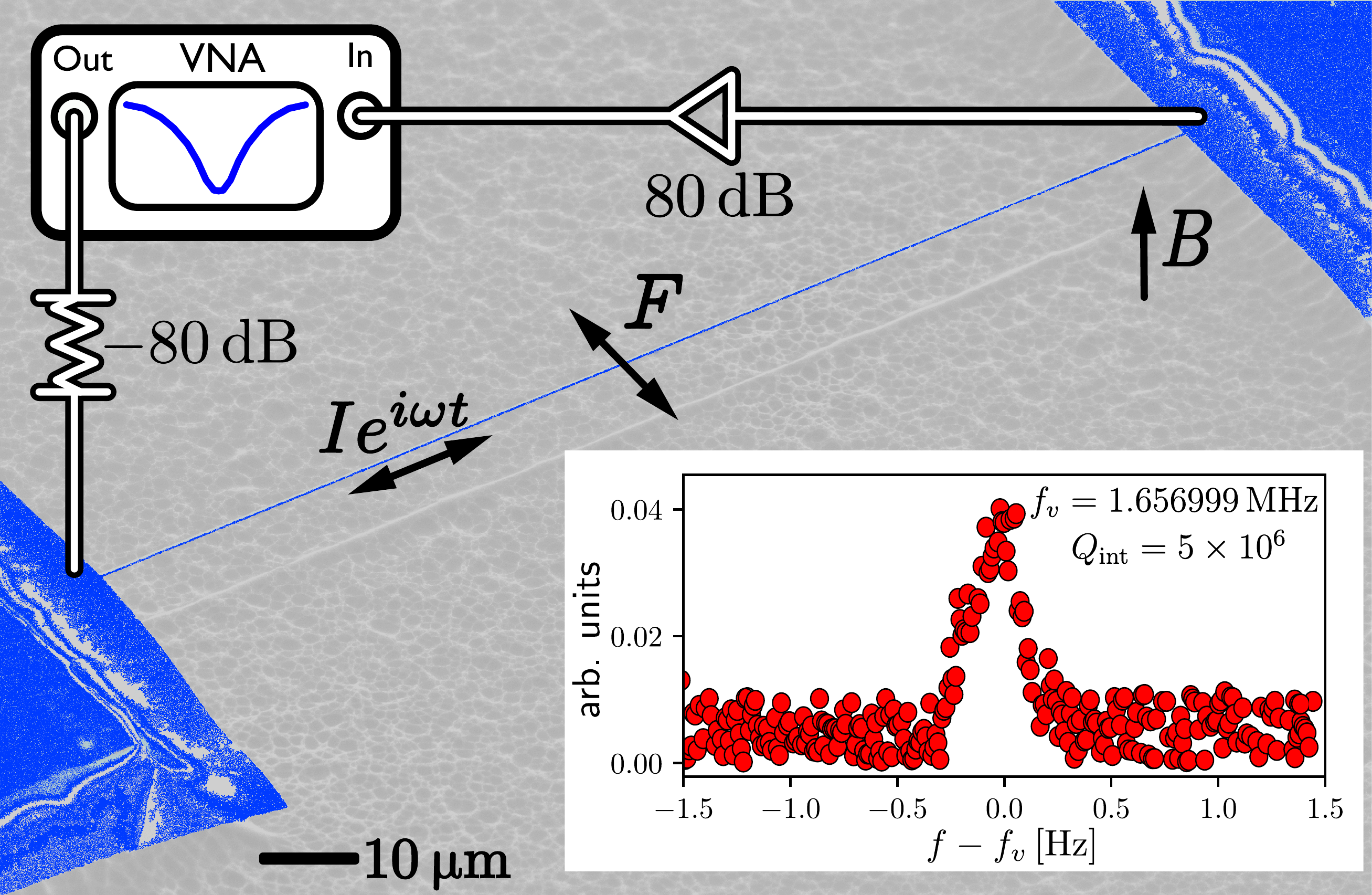}
	    \caption{\label{setup} (Color Online) False color scanning electron microscope image of a \(150\,\mathrm{\mu m}\) long composite aluminum on silicon nitride nanobeam. The device is driven by a network analyzer through \(80\,\mathrm{dB}\) of attenuation distributed over several temperature stages of the cryostat. At the output, two \(40\,\mathrm{dB}\) low-noise amplifiers are used at room temperature to improve the signal-to-noise ratio. Inset shows an example of a frequency response taken in vacuum at \(7\,\mathrm{mK}\) with a magnetic field of \(10\,\mathrm{mT}\) demonstrating a quality factor of \(5\times10^6\).}
	\end{figure}

	We present operation of two magnetomotively driven, doubly-clamped, composite nanobeams consisting of an aluminum film deposited on pre-stressed silicon nitride from \(4.2\,\mathrm{K}\) down to \(7\,\mathrm{mK}\) in liquid \(^4\mathrm{He}\). Superfluid \(^4\mathrm{He}\) is an ideal starting place for studying the interaction of nanoscale mechanical objects with quantum fluids,  since the viscosity of the \(^4\mathrm{He}\) is lower than that of \(^3\mathrm{He}\)  and the superfluid transition temperature is higher \(\sim 2.17\,\mathrm{K}\). Using the well understood range of thermal excitations \cite{hohenberg2000microscopic} and topological defects \cite{tsubota2009quantum} of superfluid \(^4\mathrm{He}\) \cite{enss2005low} we directly contrast established theoretical framework with the experimental observations. In our earlier work, we have already demonstrated that the nanobeams are extremely sensitive to the normal-fluid component in the hydrodynamic regime \cite{bradley2017operating}, above \(1\,\mathrm{K}\). Here we extend this research to the ballistic regime, in the temperature range down to \(7\,\mathrm{mK}\), detecting the presence of the thermal excitations: phonons and rotons. Furthermore, we argue that nanobeam motion in superfluid generates acoustic waves (first sound), a mechanism that dominates damping at the lowest temperatures. 
	
	\begin{figure*}
		\includegraphics[width=0.98\linewidth]{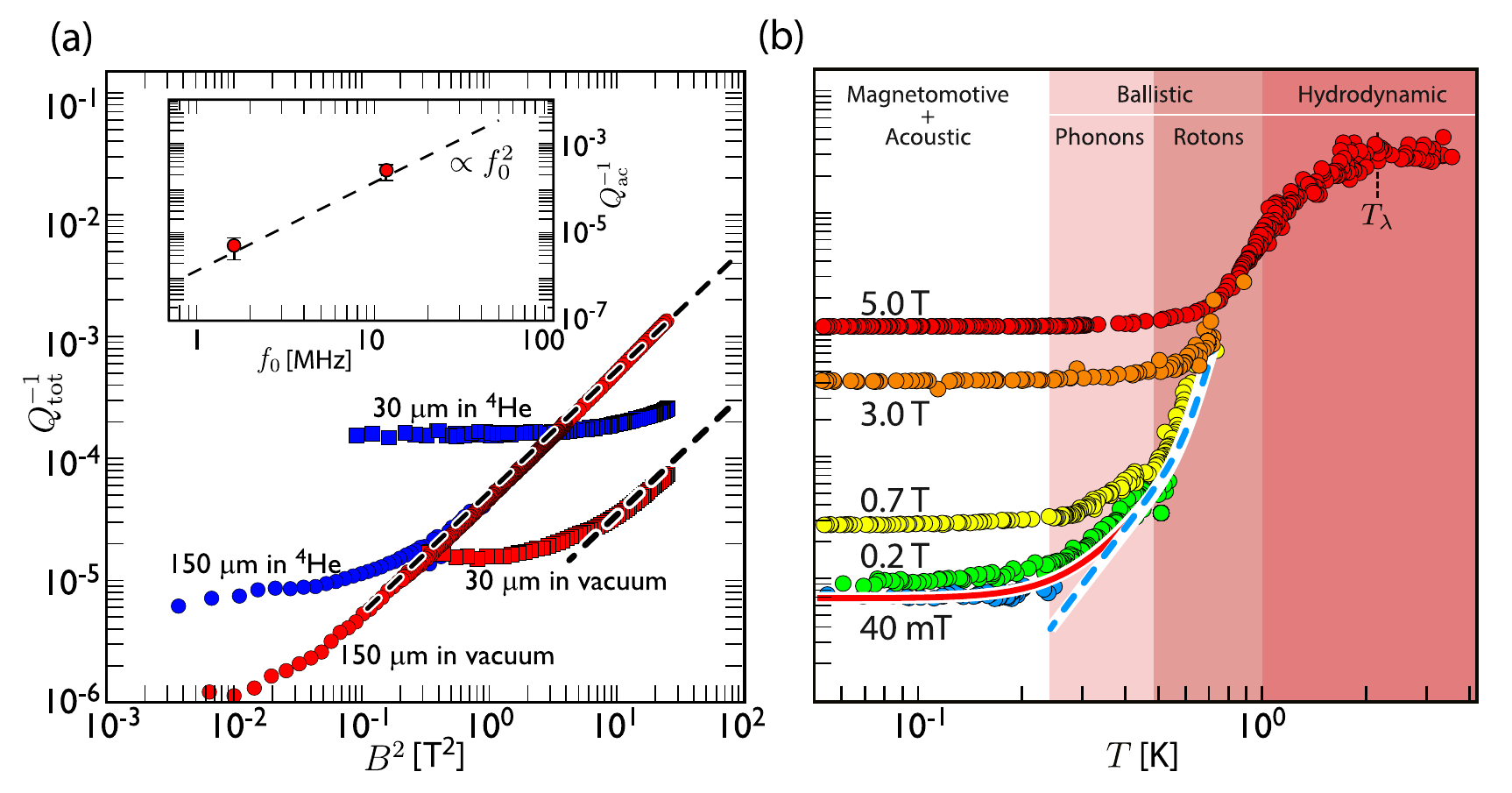}
		\caption{\label{Fig:FieldDepTempDep}(a) (Color Online) Magnetic field dependence of damping for \(150\,\mathrm{\mu m}\) and \(30\,\mathrm{\mu m}\)-long beams operated in vacuum and superfluid \(^4\mathrm{He}\) at \(10\,\mathrm{mK}\). Damping at high magnetic fields follows \(B^2\) dependence (shown with dashed lines) as a result of magnetomotive loading in nanobeams \cite{voncken1998vibrating}. Towards low magnetic fields \(Q^{-1}_{\mathrm{tot}}\) becomes constant as we approach the internal damping of the beams in vacuum along with acoustic emission in the liquid. Inset: Frequency dependence for acoustic damping in superfluid \(^4\mathrm{He}\). The dashed line shows the \(f_0^2\) dependence calculated by Eq.\,(\ref{acoustic1}) for dipole acoustic emission with no fitting parameters \cite{landau1987fluids}. (b) Temperature dependence of damping for a \(150\,\mathrm{\mu m}\)-long sample immersed in a liquid \(^4\mathrm{He}\) measured in a range of magnetic fields. Notations on the top indicate dominant damping mechanisms for each temperature range, with \(T_{\lambda} = 2.17\,\mathrm{K}\) indicating the superfluid transition temperature. The damping of the nanobeam below \(1\,\mathrm{K}\) is explained by ballistic scattering of the thermal excitations in superfluid \(^4\mathrm{He}\) on the nanobeam surface. The combined damping due to phonons Eq.\,(\ref{eq:tempdep1}) and rotons Eq.\,(\ref{eq:tempdep2}) is shown by the blue dashed line. The red solid line shows total damping calculated as a sum of phonon, roton and acoustic emission term Eq.\,(\(\ref{acoustic1}\)). For magnetic fields exceeding \(0.1\,\mathrm{T}\) the magnetomotive damping becomes dominant at low temperatures.}
	\end{figure*}

	The two devices discussed in this work have width \(w=300\,\mathrm{nm}\), thickness \(t=130\,\mathrm{nm}\), and length \(l=30\,\mathrm{\mu m}\) and \(150\,\mathrm{\mu m}\), with fundamental mode frequencies of \(11.6\,\mathrm{MHz}\) and \(1.6\,\mathrm{MHz}\), respectively. We fabricated devices on commercially available, undoped silicon wafers covered with a \(100\,\mathrm{nm}\) thick layer of silicon nitride, pre-stressed to $0.9 \, \mathrm{GPa}$. The deposited \(30\,\mathrm{nm}\) thick aluminum layer was patterned using electron-beam lithography and dry-etched in chlorine-based plasma. Suspension of the doubly clamped beams was performed by creating an undercut through selectively etching Si in \(\mathrm{XeF_2}\). Figure\,\ref{setup} shows an SEM image of the longest \(150\,\mathrm{\mu m}\) nanobeam and principle measurement scheme. For the experiment the nanobeams were driven using an AC current that passed through the devices in a constant, perpendicular magnetic field, \(B\). The resulting Lorentz force induced oscillatory motion of the samples, creating a back-emf which was measured as a drop in the transmitted signal \(S_{21}\) of a vector network analyzer.

	The measurements presented here were taken in a brass cell inside a cryogen-free dilution refrigerator with a base temperature of \(7\,\mathrm{mK}\). The cell was attached to the mixing chamber plate and filled with \(^4\mathrm{He}\) using two capillary lines, which were thermally anchored to each temperature stage of the refrigerator using silver-sintering. The cell temperature was inferred from a calibrated \(\mathrm{RuO_2}\) thermometer thermally anchored to the mixing chamber. 

	Initially, we characterized the nanobeams in vacuum to determine the internal losses for the two resonators, both in the normal and superconducting states of the aluminum. The inset in Fig.\,\ref{setup} shows the resonance response in vacuum for the \(150\,\mathrm{\mu m}\) beam in the superconducting state at base temperature in a \(10\,\mathrm{mT}\) magnetic field.  The measurements show that under these conditions the internal losses of the nanobeam, defined as the inverse quality factor, are very low, \(Q_\mathrm{int}^{-1}\approx2\times 10^{-7}\). These losses are presumably caused by an ensemble of two-level fluctuators, as has been previously demonstrated in experiments involving superconducting resonators \cite{lulla2013evidence}, since the clamping losses are expected to be extremely small due to the high aspect ratio of the resonators \cite{wilson2008instrinsic}.

	Figure\,\ref{Fig:FieldDepTempDep}(a) shows the dependence of damping of the nanobeams on the external magnetic field from \(60\,\mathrm{mT}\) up to \(5\, \mathrm{T}\). In these magnetic fields aluminum is in the normal state, and internal damping of the \(150\,\mathrm{\mu m}\) beam is \(\sim10^{-6}\). This damping is magnetic field independent up to \(150\,\mathrm{mT}\), above which the magnetomotive loading (dissipation arising from moving the conductor in the magnetic field) starts to dominate, with the expected \(B^2\) dependence. The behavior of the \(30\,\mathrm{\mu m}\)-long beam in the magnetic fields is similar but the internal losses are larger and hence remain field-independent up to \(1\,\mathrm{T}\).  

	After characterization in vacuum, we operated the nanobeams in liquid \(^4\mathrm{He}\) over the temperature range from \(7\,\mathrm{mK}\) to \(\sim3\,\mathrm{K}\). This is not trivial, since over this interval the damping arising from the surrounding liquid helium varies over four orders of magnitude. We therefore had to adjust the magnetomotive force according to the level of damping, meaning that measurements had to be made over a range of magnetic fields up to \(5\,\mathrm{T}\).

	Figure\,\ref{Fig:FieldDepTempDep}(a) shows that both the \(150\,\mathrm{\mu m}\) and \(30\,\mathrm{\mu m}\)-long beams demonstrated an order of magnitude larger damping in superfluid \(^4\mathrm{He}\) at \(7\,\mathrm{mK}\) than in vacuum, even in low magnetic fields where magnetomotive losses are negligible. This is rather counter-intuitive: the Stokes' drag from interaction with the superfluid normal component should be negligible at these temperatures, since the density of thermally excited phonons and rotons is insignificant \cite{morishita1989mean}. Later in this paper we will show that these losses can be attributed to acoustic emission in superfluid \(^4\mathrm{He}\). Since the behavior of the \(30\,\mathrm{\mu m}\)-long beam in the fluid is almost independent of the magnetic field, we have mainly focused our studies on the \(150\,\mathrm{\mu m}\)-long beam. 

	In order to clarify the damping mechanisms in the liquid helium, the temperature dependence of the quality factor of the \(150\,\mathrm{\mu m}\)-long beam was measured, as shown in Fig.\,\ref{Fig:FieldDepTempDep}(b). The measurements map out rather consistently the damping arising from the liquid over the whole temperature range. As can be seen, in addition to the beam's internal, \(Q_\mathrm{int}^{-1}\), and magnetomotive, \(Q_\mathrm{mm}^{-1}\), losses, there are three distinct temperature regimes distinguished by the dominant damping mechanisms originating in the liquid. These are: (with increasing temperature) acoustic emission \(Q_\mathrm{ac}^{-1}\), ballistic scattering of thermal excitations (phonons, \(Q_\mathrm{ph}^{-1}\), and rotons, \(Q_\mathrm{rot}^{-1}\)) and finally, above \(1\,\mathrm{K}\), hydrodynamic losses \(Q_\mathrm{hy}^{-1}\). These mechanisms together constitute the total damping:
	\begin{equation}
		Q_\mathrm{tot}^{-1}= Q_\mathrm{int}^{-1} + Q_\mathrm{mm}^{-1}+ Q^{-1}_\mathrm{ac}+Q_\mathrm{ph}^{-1}+Q_\mathrm{rot}^{-1}+Q_\mathrm{hy}^{-1}.
		\label{Eq.Sum}
	\end{equation}
	It is remarkable that all these distinct behaviors can be extracted with a single device. The hydrodynamic Stokes' drag by the normal fluid component is substantial at temperatures above \(1\,\mathrm{K}\) and can be described in the framework of the phenomenological two-fluid model of superfluid \(^4\mathrm{He}\) \cite{bradley2017operating}.

	To model the damping at lower temperatures we can take the dissipation as arising from the scattering of individual phonons and rotons at the beam surface. This is justified by the fact that below \(\mathrm{1\,K}\) the mean free paths of thermal excitations in the superfluid (rotons and phonons) are longer than the effective nanobeam diameter, while the de Broglie wavelengths are much shorter \cite{morishita1989mean}. The phonon drag force, $F_{\mathrm{ph}}=-A \rho_{\mathrm{ph}} c_{\mathrm{ph}} v_w d/2$, exerted per unit length of the beam is obtained by summing the momentum of phonons colliding against the beam, with phonon density, $\rho_{\mathrm{ph}}$, effective beam diameter, \(d\approx150\,\mathrm{nm}\), phonon velocity, \(c_\mathrm{ph}=229\,\mathrm{m\,s^{-1}}\), and wire velocity, $v_w$. This gives the damping contribution \cite{morishita1989mean}:
	\begin{equation}
		Q_\mathrm{ph}^{-1} = \mathcal{A}\frac{k_\mathrm{B}^4}{45 \hbar^3 df_0(\rho + \rho_s) c_\mathrm{ph}^4} T^4,
		\label{eq:tempdep1}
	\end{equation}
	 where \(f_0\) is the resonance frequency of the beam in liquid, \(\rho\) and \(\rho_s\) are the densities of the beam material and superfluid respectively, and \(\mathcal{A}=2.67\) is a constant taking into account the ``cylindrical'' beam geometry \cite{morishita1989mean}.

Following similar arguments, however now using the number density of rotons, $n_\mathrm{rot}$ \cite{landau1965superfluidity}, the roton drag force is $F_\mathrm{rot} = -A n_\mathrm{rot} p_0 v_w d/2$, and the damping arising from roton scattering becomes:

	\begin{equation}
		\label{eq:tempdep2}
Q_\mathrm{rot}^{-1} = \frac{\mathcal{A} p_0}{\pi^2 d f_0 \rho_w }\left( \frac{m^* k_\mathrm{B} T}{2 \pi \hbar^2} \right)^{3/2} \exp\left(-\frac{\Delta}{k_\mathrm{B} T}\right),
	\end{equation}
	where \(p_0\), \(m^*\) and \(\Delta\) are the Landau roton parameters \cite{morishita1989mean}.


	The combined damping from the phonon and roton scattering, \(Q_\mathrm{ph}^{-1} + Q_\mathrm{rot}^{-1}\), is shown by the blue dashed line in Fig.\,\ref{Fig:FieldDepTempDep}(b), and describes well the envelope of the experimental data in the temperature range \(0.4\,\mathrm{K}-0.7\,\mathrm{K}\) with no fitting parameters. At lower temperatures, the deviation from the experimental results and theoretical predictions is clear, and measured damping saturates at \(\mathrm{mK}\) temperatures. 

	This implies the presence of another temperature independent damping mechanism, which is significantly larger than that of the intrinsic losses of the beams measured in vacuum. Figure\,\ref{Fig:FieldDepTempDep}(a) shows that damping in the liquid reaches \(Q_\mathrm{ac}^{-1}\approx5\times10^{-6}\) for the \(150\,\mathrm{\mu m}\)-long, and \(Q_\mathrm{ac}^{-1}\approx2\times10^{-4}\) for the \(30\,\mathrm{\mu m}\)-long nanobeams, respectively. We attribute this mechanism to acoustic losses which is described in the framework of dipole emission \cite{landau1987fluids}:
	\begin{equation}
		\label{acoustic1}
		Q_\mathrm{ac}^{-1} = \frac{\pi^3}{2}\frac{\rho_s}{\rho}\left(\frac{ df_0}{c_\mathrm{ph}}\right)^2,
	\end{equation}
	predicting a quadratic dependence on the frequency of the nanobeam oscillating in fluid. The inset in Fig.\,\ref{Fig:FieldDepTempDep}(a) shows the measured losses of the nanobeams in superfluid helium as a function of their resonance frequencies, and contrasts them with the expected damping from acoustic emission obtained from Eq.\,(\ref{acoustic1}) without fitting parameters. Remarkably, the measured values of losses for both beams are in excellent agreement with this dependence.

	The solid line in Fig.\,\ref{Fig:FieldDepTempDep}(b) shows the resulting sum of the damping mechanisms from the thermal excitations in the helium and acoustic emission from the nanobeam. This dependence describes well the data for the lowest magnetic field where the magnetomotive damping is negligible. Existence of considerable acoustic emission at high frequencies, should be taken into account for the development of the phononic experiments in quantum fluids\cite{schwab2005putting, fong2018phonon}. According to Eq.\,\ref{acoustic1} acoustic losses are expected to be negligible at lower frequencies, where the condensate should behave essentially as a mechanical vacuum.	

	The acoustic losses could be reduced by utilizing lower frequency detectors, or beams with a smaller diameter, for example, carbon nanotubes. Alternatively, the mechanical resonators can be enclosed in a cavity, with a size smaller than half of the acoustic wavelength of the emitted sound along the axis of the dipole emission. This will restrict the number of acoustic modes available for emission. For the frequency of \(\sim1.6\,\mathrm{MHz}\), the characteristic size of the acoustic cavity should be \(\sim50\,\mathrm{\mu m}\). Nanobeams could also be utilized for studies in thin-film rather than bulk superfluid, as has been previously demonstrated using nanotubes \cite{noury2019layering}.
	
	In summary, nanobeams have been measured in \(^4\mathrm{He}\) down to \(\mathrm{mK}\) temperatures, characterizing the various dissipation mechanisms. We observe that ballistic quasiparticles dominate damping below \(1\,\mathrm{K}\), with acoustic emission taking over at the lowest temperatures. We demonstrate excellent agreement with a model incorporating acoustic emission, and interaction with ballistic phonons and rotons in the low-temperature limit. Versatility of the devices could be further improved by suppressing acoustic losses through mounting the nanobeams in some confined enclosure. Nanobeams provide a convenient, tunable and precise tool to characterize the properties of an environment with high accuracy. This experiment paves the way towards using nanobeams as ultrasensitive detectors for probing the properties of \(^3\mathrm{He-B}\) at \(\mathrm{\mu K}\) temperatures, where the size of the probe becomes comparable to the superfluid coherence length. 
	
	All data used in this paper are available at http://dx.doi.org/10.17635/lancaster/researchdata/xxxx, including descriptions of the data sets. We would like to acknowledge the excellent technical support of A.~Stokes, M.\,G.~Ward, M.~Poole and R.~Schanen and very useful scientific discussions with S.~Autti, E.~Laird, A.~Jennings, M.\,T.~Noble and T.~Wilcox. This research was supported by the UK EPSRC Grants No. EP/L000016/1, EP/P024203/1 and No. EP/I028285/1, and the European Microkelvin Platform, ERC 824109. EC acknowledges the support from the ANR grant MajoranaPRO No. ANR-13-BS04-0009-01 and the ERC CoG grant ULT-NEMS No. 647917.

\end{document}